\documentclass[a4papper,oneside,notitlepage,8pt,twocolumn]{amsart}
\usepackage{amsaddr}
\usepackage[T1]{fontenc}
\usepackage[english]{babel}
\usepackage[a4paper]{geometry}

\usepackage[cm]{fullpage}
\usepackage{xspace}
\usepackage{amsmath,mathtools}
\usepackage[]{kpfonts}
\usepackage{xfrac}
\usepackage{urwchancal}
\usepackage{wrapfig}
\usepackage[framemethod=TikZ]{mdframed}
\DeclareMathAlphabet{\mathpzc}{OT1}{pzc}{m}{it}
\usepackage{siunitx}
\usepackage{esint}
\usepackage{hyperref}

\newcommand{\sech}{\mathrm{sech}}
\newcommand{\ud}{\mathrm{d}}
\newcommand{\br}[1]{\left(#1\right)}
\newcommand{\ef}{\mathfrak{e}}
\newcommand{\mf}{\mathfrak{m}}
\newcommand{\micro}{\textnormal{\textmu}}
\newcommand{\muGs}{\,\micro{\rm Gs}}
\newcommand{\Gs}{\,{\rm Gs}}
\newcommand{\eV}{\,{\rm eV}}
\newcommand{\GeV}{\,{\rm GeV}}
\newcommand{\TeV}{\,{\rm TeV}}
\newcommand{\I}{\mathrm{i}}

\begin{document}

\title{On the issue of magnetic monopoles in the prospect of UHE photon searches.}
       
\author{{\L}ukasz Bratek \  and \  Joanna Ja{\l}ocha}
\address{Institute of Physics, Cracow University of Technology,  
ul. Podchor{\k a}{\.z}ych, PL-30084 Krak{\'o}w, Poland}
\email{Lukasz.Bratek@pk.edu.pl}

\date{\today}

\begin{abstract}
Ultra-high energy (UHE) photons  with energies exceeding $10^{18}\eV$ can potentially be observed. They are produced in various processes involving electrically charged particles.  
However, more exotic scenarios are also possible.
UHE photons could be emitted in encounters of massive magnetically charged monopole-- antimonopole pairs or in  processes associated with monopoles accelerated to high energies, typically $10^{21}\eV$ or beyond. Observing UHE photons can pose constraints on the properties of magnetic monopoles.
 There are compelling theoretical reasons in favor of the presence of magnetic monopoles in nature.    
The predicted observational signatures of these particles are therefore searched for in dedicated experiments currently in operation. Despite these attempts, magnetic monopoles have yet to be empirically proved.
There are also theoretical reasons why {\bf magnetic monopoles allowed by Dirac's theory 
might not be realized in nature in the form of isolated particles.} 
Detection or non-detection of UHE photon signatures of magnetic monopoles would bring us closer to solving this fascinating puzzle.
\bigskip\newline  {\bf keywords:} magnetic monopole; zero frequency fields; quantum theory of the electric charge; 
charge quantization, fine structure constant, ultra high energy photons, existence of magnetic monopoles
\end{abstract}  

\maketitle

\section{Introduction}

The upper limit on the distance from which high-energy cosmic rays can travel is one of the key predictions in cosmic ray physics \cite{1_greisen1966,2_zatsepin1966}. 
Interactions with background radiation result in significant energy loss for protons with energies greater than  $5{\times}10^{19}\eV$. These particles are expected to travel only a few mega-parsecs before their energy falls below the threshold for photoproduction. Similarly, if the cosmic ray particle is a heavier ion, the interaction with background photons reduces its energy. Therefore, a truncation in the cosmic ray spectrum known as the GZK limit is expected. Cosmic rays with energies above $10^{20}\eV$,  coming from sources at distances greater than a few dozens of mega-parsecs, should be invisible due to the GZK limit \cite{3_cronin1992,4_yoshida1993}.
The truncation of the cosmic ray spectrum can also be a result of the properties of the sources themselves, not just due to the GZK effect. It cannot be excluded that there are no mechanisms accelerating particles above  $10^{20}\eV$.
If the extension of detection capabilities would result in an increase in the observed flux of cosmic radiation at the highest energies, this would then require
consideration of particle acceleration scenarios, which currently seem to be rather exotic.

Among possible sources of cosmic radiation of the highest energies, most promising are those in the presence of very strong magnetic fields extending over long distances.  Second-order Fermi acceleration processes \cite{fermi1949} occur in the presence of randomly moving plasma clouds carrying magnetic fields through the interstellar medium. An electrically charged particle can gain energy by repeatedly bouncing off plasma clouds.  However, there is a limit to the efficiency of the acceleration process. The particle must stay in a magnetic field region for the acceleration to occur: if the particle's energy increases too much, the particle will escape from that region. The maximum energy gained is $E_F{\approx} Z \beta c e B L$, where $\beta c$ is the magnetic mirror's velocity, $B$ is the magnetic field strength, and $L$ is the size of the acceleration region. Particles can also be accelerated in shock fronts, bouncing back and forth across the shock, and if shock waves are also present, the acceleration process will be more efficient \cite{bell1978}. It was also shown  \cite{gallant1999,blasi2000} that it is possible to accelerate particles in the vicinity of pulsars even to energies of $10^{21}\eV$. This is a debatable result.
However, if the accelerated particle is magnetically charged, the situation changes radically. 
Magnetic monopoles could explain the production of particles of the highest energy even in the region of our Galaxy.

As suggested in 1960 \cite{porter1960}, a number of effects observed  in air showers could be understood if a fraction
(about $10^{-14}$) of all primary cosmic ray
particles were in form of magnetic monopoles, at that time assumed to be 
point-like elementary magnetic charges predicted by Dirac in 1931 \cite{dirac1931}.
The monopole idea is unavoidable when considering an electrically charged quantum particle coupled to genuine Maxwell fields.  There is an unobservable singularity string emanating from the monopole position-point.  The monopole mass remains unspecified in Dirac's theory. 
It was later discovered that a magnetic monopole can be realized in terms of field-theoretical concepts. 
Topological magnetic monopoles emerge in unified non-Abelian gauge field theories of interactions via a mechanism similar to that of a monopole discovered independently by 't~Hooft \cite{hooft1974} and Polyakov \cite{polyakov1974} inan  $SO(3)$ realization of the Georgi--Glashow model \cite{georgi1972}. The monopoles are spherically symmetric massive spatially extended solitonic solutions without a singularity string.  The solitons are characterized by a topological charge -- an integer describing the winding degree of a finite energy mapping between the physical space domain and the field configuration domain. In the low energy regime of Unified Theories of Interactions, the topological charge can be reinterpreted in terms of magnetic charge of a $U(1)$ gauge potential equivalent to one of a genuine Maxwell theory. The resulting magnetic field is asymptotically equivalent to 
that of Dirac's monopole.
The problem of the relationship of magnetic monopoles to cosmic radiation of the highest energies consists essentially of two separate questions. The first question is more experimental: {could magnetic monopoles be responsible for the production of UHE photons in the range of highest energies observed for cosmic rays?} The second question is more theoretical: { do magnetic monopoles exist?}

The acceleration of monopoles in magnetic fields would happen directly and  efficiently; no Fermi-type processes would be needed.
Magnetic monopoles, so far hypothetical particles, could form in the dense matter of neutron stars \cite{milton2006}.
Magnetic monopoles can be produced and efficiently accelerated by neutron stars' magnetic fields (especially if they contain quark matter). Magnetic monopoles may also play a role in various astrophysical processes, such as mechanisms that reduce magnetic fields in the interiors of neutron stars \cite{harvey1984}.
  Magnetic monopoles could be produced in strong magnetic fields via mechanism analogous to 
Schwinger pair production mechanism in which electrically charged particles are created by tunelling through a barrier in extremely strong electric field \cite{schwinger1951}. By duality of electromagnetic equations, magnetically charged particles are also expected to be produced by tunneling in sufficiently strong magnetic fields.  This expectation is confirmed by rigor semi-classical calculations of the rate of monopole pair production in a constant magnetic field. In the case of point-like monopoles \cite{affleck1982}, the resulting production rate bears resemblance to that of Schwinger for electron-positron pair production, at least in the leading term. The rate of production in the case of non-point-like  't~Hooft and Polyakov monopole  was calculated in \cite{ho2021}. It turns out that for heavy monopoles the required magnetic field for this process would exceed those in neutron stars. However, extremely strong electromagnetic fields can be generated in high energy heavy-ion collisions in which nuclei are accelerated to nearly the speed of light. Magnetic fields are induced by electric currents of positively charged nuclei and are the largest component of the electromagnetic field near vicinity of the collision center. In Earth-based laboratories one can achieve energies of $200\GeV$ for Au+Au collisions (RHIC) and $2.76\TeV$ for Pb+Pb collisions (LHC), giving rise to huge magnetic fields of $10^{18}$--$10^{19}\Gs$ and $10^{20}\Gs$, respectively,  as calculated in \cite{huang2016},  exceeding highest values known for astrophysical sources, such as  $10^{15}\Gs$ on the surfaces of magnetars \cite{magnetars2017}. 
Recently, presented was the first search for finite-size monopoles created via the Schwinger mechanism.   The search was conducted by the MoEDAL experiment, designed specifically to see monopoles
directly. The experiment is observing Pb-Pb collisions producing strong magnetic field in which monopole-antimonopole pairs may be created via tunneling effect. The analysis of the data based on a nonperturbative cross-section calculations allowed to place the (conservative) lower mass  limit of $75\GeV$ for magnetic charges between $1$ and $3$ Dirac charges produced via Schwinger mechanism
\cite{acharya2022}.
 It is clear that magnetic monopole signatures have the potential to be detected and should be sought for.
Several other experiments have been carried out in the search for magnetic monopoles. Various upper limits have been established so far: MACRO \cite{ambrosio2002},  Baikal neutrino
  telescope \cite{baikal2005}, 
  Amanda-II Detector \cite{wissing2007},  RICE \cite{rice2008},
   SLIM \cite{slim2008}, ANITA II interferometer \cite{anita2011},
   IceCube detector \cite{icecube2013} and Pierre Auger Observatory \cite{fujii2016}.
The current state of magnetic monopole searches is discussed in \cite{giacomelli2005}.  The discussion on the theoretical and experimental status of magnetic monopoles and the limits on monopole masses can be found in \cite{milton2006}.    Recent advances in the theoretical and experimental physics of magnetic monopoles are discussed in \cite{mavromatos2020}. A comprehensive review on various aspects of the theory of magnetic monopoles can be found in \cite{shnir2005}.

In the context of UHE photons, it is important to note that processes characterized by photon emission from magnetic monopoles with the lowest nonzero magnetic charge allowed by Dirac theory  \cite{dirac1931} will be enhanced by a factor of $4692$ over similar processes involving unit electronic charge  \cite{dooher1971}.  
Magnetic charges of monopoles realized in nature can be  $n$ times greater than the allowable minimum, where $n$ is an integer; hence the number $4692$ can be still increased by a factor $n^2$
(for example, $n{=}2$ in a Schwinger field theoretical model of monopoles \cite{schwinger1966}).
In effect, one may expect radiation enhanced by many orders of magnitude  in processes in which magnetic monopoles take part.
The highly increased electromagnetic coupling with matter also implies strong ionizing properties of magnetic monopoles, which is  experimentally advantageous.   Magnetic monopoles or their electrically charged counterparts dyons are hoped to be found among highly ionizing particles in the MoEDAL search experiment at  the LHC \cite{mitsou2022}. 

It seems, however, that the 
highest energy scale of $10^{21}\eV$ or more, exceeding the highest 
energies of $10^{20}\eV$ observed so far for UHECR \cite{aab2020}, can be ascribed to the interaction of magnetic monopoles with
magnetic fields in the astrophysical context in a universal way, independently of the monopole mass (this issue will be discussed later). 
The possibility to accelerate the monopole to extreme energies \cite{porter1960,kephart1996} or to have monopolia  \cite{hill1983} present in nature is very important in the context of detection of UHE photons. If an already accelerated high energy monopole hits the regions where it interacts with matter then there will be emission of photons. 
Could this energy be then deposited in several or even two single UHE photons? The answer to this question seems affirmative,  as discussed later.   The possibility to detect photons in cosmic radiation would be very promising. If monopoles exist and are accelerated to decay, such high-energy photons should be a component of cosmic radiation. However, a photon, through processes of pair creation, can ultimately also give rise to charged particles of cosmic radiation of the highest energy.  Observing such photons can pose constraints on the properties of magnetic monopoles.   UHE photons with energies above $10^{19}\eV$  can be detected
with the Surface Detector array of the Pierre Auger Observatory \cite{bleve2015}. 
As suggested in \cite{epele2008,epele2009}, monopolia might be easier to detect than free monopoles
and could be discovered at the LHC. In this context photon production by monopole loops in photon fusion was studied in \cite{epele2012}, in particular,
the cross-sections for the annihilation of monopoles into two photons were calculated and discussed. Such loops may manifest themselves in light-by-light scattering at the LHC \cite{ellis2017} or future colliders \cite{ellis2022}. Due to the very strong coupling between monopoles, monopole-antimonopole pairs (forming monopolium neutral bound states rather than unbound free states due to the strong coupling) may finally annihilate into highly energetic photons and thus be observed.
  
Before making any claims regarding the presence or absence of magnetic monopoles in nature, it would be appropriate to note that this term refers to two classes of objects.
 The first class consists of elementary point-like magnetic monopoles described by Dirac \cite{dirac1931}, resembling the Dirac electron more in structure than a composite particle. This is an ideal mathematical construct that gives
the reason for why any magnetic monopole that could be realized as a physically existing particle must have its magnetic charge as an integer multiple of some 
 elementary magnetic charge $\frac{e}{2\alpha}$ (here, $e$ is the electronic charge and $\alpha$ is the fine structure constant). 
The second class consists of  particular realizations of the ideal elementary magnetic monopole within a field-theoretical framework characteristic of the  standard model of particles or other theories of  unified interactions. These solitonic-like particles resemble more the composite proton with a complicated internal structure rather than the elementary electron.   The classic example is provided by the magnetic monopole of 't~Hooft and Polyakov \cite{hooft1974,polyakov1974}. This model will be shortly discussed in order to see the conceptual difference between the elementary monopole and its physical realizations. 

In the context of UHE photons produced in processes involving magnetic monopoles, the present work addresses the theoretical question of the existence of a magnetic monopole as an isolated particle with a non-zero multiple of elementary magnetic charge. It is enough to focus on the elementary 
 Dirac's magnetic monopole for this purpose. The problem of a particular field-theoretical realization of the magnetic monopole is different. Although isolated magnetically charged solitonic configurations can be considered as solutions in the standard model of particles, it might be that only magnetically neutral configurations of magnetic monopoles are realized in nature  (similarly , only color-neutral quark configurations are observed in the form of isolated particles). 
  
    The arguments  against the existence of magnetic monopoles are formulated within the framework of the asymptotic (infrared) structure of the Maxwell theory and the principles of quantum theory
 (unfortunately, this regime of electromagnetic field presents some theoretical difficulties \cite{gervais1980,herdegen2012}).  
In Herdegen's argument \cite{herdegen1993}, the absence of magnetic charges serves as the consistency condition for the possible unambiguous extension of the definition of angular momentum to the typical situation where massive charged particles or fields (such as described by Dirac or Klein Gordon equations) are scattered from initial to final free asymptotic states. This argument will not be discussed here.
The present paper focuses on Staruszkiewicz's argument against the existence of magnetic monopoles \cite{AStar1998a}.
 This argument is an important part of quantum theory of the electric charge \cite{AStar1989a}. This is an `emergent' theory, separated from Quantum Electrodynamics by a nontrivial limit $r{\to}\infty$.
 Except for explaining the
 quantization of electric charge in terms of a single universal constant \cite{AStar1998c} (which is done independently of the concept of magnetic monopole), the theory has a nontrivial dependence on the numerical value of the fine structure constant $\alpha{=}\frac{e^2}{\hbar c}$. In particular, this theory leads to mathematically critical spectra of $\alpha$ values (however, it is not yet known if this theory also predicts the experimentally determined value of $\alpha$). In Dirac's theory \cite{dirac1931} the very presence of a single magnetic monopole leads to quantization of electric charge; however, the value of the fine structure constant remains an arbitrary parameter. In Dirac' theory, the condition for quantization of electric charge is topological in character, whereas in Staruszkiewicz's theory the quantization of electric charge arises from a quantal eigenvalue problem.

\section{Magnetic monopoles and UHE photons} 
 
In classical electrodynamics of Maxwell based on a single gauge potential four-vector, there is no place for magnetic monopoles, and the reason for this is 
structural. Introducing magnetic monopoles is, nevertheless, still possible with only a slight modification 
of the theory to account for topological changes introduced by considering quantum charged systems. 
The Dirac monopole theory is introduced in the framework of ordinary Maxwell theory where there is 
an analytical global correspondence in spacetime between a gauge-invariant field strength bivector $F$
and a single gauge potential $A$. Considering the magnetic monopole within the Maxwell theory framework would be possible only by allowing for braking this correspondence locally. In quantum mechanics, the phase of quantum states is unobservable. This fact, in conjunction with Maxwell theory, led Dirac to discover that certain singularities in phase can manifest themselves as point-like magnetic monopoles. The monopoles can be consistently considered on theoretical grounds, since arbitrariness in quantum phase can be absorbed by the arbitrariness of gauge potential of the electromagnetic field. This issue is addressed later in more detail.

\subsection{Accelerated monopoles and a coincidence with the highest energies observed for 
cosmic rays}

In the sequel to his paper on magnetic monopole \cite{dirac1931}, Dirac investigated the general problem of motion
of electric and magnetic charges interacting electromagnetically with each other \cite{dirac1948}.
Dirac assumed that the motion of a point magnetic charge $g$ of mass $m$ in the electromagnetic field $F_{\mu\nu}$
is described by an equation quite analogous to Lorentz's equation $m\ddot{x}_{\mu}{=}e\dot{x}^{\nu}F_{\mu\nu}(x)$ for the motion of 
a point electric charge $e$ of mass $m$ in the same field. Namely, the equation reads: $m\ddot{x}_{\mu}{=}g\dot{x}^{\nu}G_{\mu\nu}(x)$, where 
$G_{\mu\nu}{\equiv}\frac{1}{2}\epsilon_{\mu\nu}^{\phantom{\mu\nu}\alpha\beta}F_{\alpha\beta}$ is dual to
$F_{\mu\nu}$ (more explicitly, 
  $G_{01}{=}F_{23}$, $G_{23}{=}{-}F_{01}$, and the other components are obtained by cyclic permutations of indices $1,2,3$).\footnote{Here, $\epsilon$ is the completely antisymmetric
Levi--Civita pseudotensor; the indices are understood to be raised or lowered with the help of the Lorentz metric tensor; 
the dot sign denotes differentiation with respect to the proper time of a particle.} The field includes both
external fields and the particle's own field, with singularities along its worldline that have to be appropriately  avoided
in the process of solving the equations of motion. 

From the forms of the two general equations, it is evident, in particular, that
the equation of motion for a magnetic monopole in an external magnetic field is mathematically the same as 
the  equation of motion for an electric charge in an external electric field (similarly, we can expect circular motion of magnetic monopole in a uniform electric field and the associated synchrotron emission). Owing to this correspondence, we can
use known textbook formulas for the motion of an electric charge in a uniform electric field to obtain analogous results
for the motion of a magnetic charge in a uniform magnetic field. This will be useful for estimating the order of magnitude 
of energies that one can expect for magnetic charges in the astrophysical context in large-scale ordered magnetic fields. In such fields the monopoles could be accelerated to very high energies if one 
assumes that energy loss due to the interaction with the environment is not severe and can be neglected. The energy loss in such a process is indeed negligible \cite{kephart1996}. 

For the purpose of an order of magnitude estimate needed here,  it is enough to consider the relativistic accelerated motion of a magnetic monopole in a uniform magnetic field.
To put the reference scales into the astrophysical context, we may consider pulsar magnetospheres.   In a first approximation, 
the magnetic field of a magnetosphere can be described by a magnetic dipole. Owing to the size of magnetospheres,
we can assume that on a characteristic scale of $L_o{=}1{\rm\, km}$, the dipolar field lines can be considered to be roughly uniform. As for the characteristic scale of 
the magnetic field one can choose $B_o{=}10^{12}{\rm\, Gs}$, considerably lower than 
 the maximum field strengths observed in magnetospheres.
Now, using the solution to a textbook problem for relativistic accelerated motion of an electric charge in uniform electric field \cite{landau1980} and the equivalence with the motion of a magnetic monopole in uniform magnetic field, one can see see that the minimum magnetic charge predicted by Dirac's theory $g{=}\frac{e}{2\alpha}$ gets accelerated to a Lorentz factor
$\gamma{=}
1{+}\frac{e B L}{2\alpha m c}$  after it has traveled a distance $L$ in uniform magnetic field $B$. 
Multiplying $\gamma$ by the rest energy $mc^2$ of the monopole, one obtains the total energy $mc^2{+}E_{\rm acc}$, where 
 $E_{\rm acc}{=}\frac{eBLc}{2\alpha}$ is the kinetic energy acquired during the acceleration process. This can be compared with a similarly looking formula for a quite unrelated Fermi shock acceleration mechanism of first order, in which the energy gain is substantially, that is, $2\beta\alpha{\sim}\beta{/}68$ times lower ($\beta$ is the speed of the shock front in units of $c$).
 Unlike electrons, for which synchrotron radiation in a strong uniform magnetic field would be expected, the magnetic monopole can accelerate along a straight line without emission of radiation. 
   It is possible that the GZK cutoff will not be important for very massive monopoles even though the magnetic monopole should couple with matter $\frac{eg}{e^2}{=}\frac{1}{2\alpha}{\approx} 68$ times stronger than electrically charged particles  (the rest mass of the monopole is estimated to be in the range from the electro-weak scale $10^2\GeV$ up to the GUT scale $10^{16}\GeV$).    
   The simple estimation of the energy gain could also be arrived at just from dimensional analysis and may seem not a reliable one, for example, not taking into account the energy loss due to possible multiple scattering during acceleration or other radiative processes (emission of electromagnetic radiation in general accelerated motion,  e.g. synchrotron radiation in electric fields). On the other hand,  one cannot exclude that much of the energy will be nevertheless carried by the monopole during its journey to the observer. If somewhere in the Universe such conditions exist as assumed above, and at the same time there are no strong large-scale electric fields, it is possible to directly and efficiently accelerate the monopole to extreme energies.    
   
 Remarkably, such obtained energy gain $E_{\rm acc}$ is independent of the invariant mass of the accelerating monopole.    
 For the adopted conservative scales  $B_o$ and $L_o$ characteristic of pulsar magnetospheres, 
 $$E_{\rm acc}{=}\frac{eB_oL_oc}{2\alpha}{\approx} 2.1{\times} 10^{21}\eV \quad {\rm for}\quad  B_oL_o{=}10^{17}\,{\rm Gs}{\cdot}{\rm cm}{=}3.2{\times} 10^4\muGs{\cdot}{\rm pc}. $$
This is a value coinciding with the highest energies observed so far for UHECR (with the expected magnetic monopole mass of $100\GeV$ as for the electroweak unification energy scale, the corresponding Lorentz factor would be
$\gamma{\approx}2{\times}10^{10}$). To be more realistic, the product $B_oL_o$ of the characteristic scales  should be corrected by a dimensionless factor accounting for possible inhomogeneities of the field and local changes in the relative directions of magnetic field and the velocity vectors along the monopole trajectory. 
Similarly large scale of the product $ B_oL_o$ as for magnetospheres can be obtained in other astrophysical contexts. Typical for the Galaxy magnetic field of $2\muGs$ on a scale length of $100\,{\rm pc}$ gives the acceleration energy in the interstellar medium of  $1.3{\times} 10^{19}\eV$. For other example sources the estimates are the following: 
interplanetary space  ($50\muGs$, $1{\,\rm AU}$, $1.5{\times}10^{13}\eV$);
Sun-spots  ($10^3\,{\rm Gs}$, $10^{4}{\,\rm km}$, $2.1{\times}\\10^{16}\eV$);
white dwarfs  ($5{\times}10^6\,{\rm Gs}$, $10^{4}{\,\rm km}$, $10^{20}\eV$); 
radio-galaxy lobes ($10\muGs$, $10\,{\rm kpc}$, $6.3{\times} 10^{21}\eV$); 
clusters of galaxies  ($1\muGs$, $100\,{\rm kpc}$, $6.3{\times}\\10^{21}\eV$);
active galactic nuclei  ($10^4\,{\rm Gs}$, $5{\,\rm AU}$, $1.5{\times} 10^{22}\eV$); and
intergalactic medium  ($10^{-2}\muGs$, $3{\,\rm Gpc}$, $1.9{\times}10^{24}\eV$). The reference values for the $BL$ product shown here are estimated based on figure 1 in \cite{hillas1984}. Judging based on these calculations, it appears that UHE values should be commonly attainable across the Universe in processes involving magnetic monopoles. 

\subsection{Selected mechanisms of production of UHE photons associated with magnetic monopoles}

The acceleration energy $E_{\rm acc}$ discussed in the previous section could be released in various processes dual to that associated with high-energy electrically charged particles. The phenomenology is simple. The Lorentz transformation law of electric and magnetic fields implies that the magnetic field  $B$ of an ultra-relativistic monopole `induces' an electric field of magnitude $\gamma c B$ (times a  geometric factor of order unity irrelevant here) acting on electrically charged 
constituents of the medium at rest. It follows that the electromagnetic energy loss expected for a monopole with a unit magnetic charge $g$ 
should be $\frac{ge}{e^2}{=}\frac{1}{2\alpha}{\sim}68$ times greater than for an ultra-relativistic unit 
electric charge of comparable invariant mass and Lorentz factor $\gamma$ (this order of magnitude argument derives from the reasoning presented in \cite{kephart1996}).
If the accelerated magnetic monopole were to decay or annihilate with another anti-partner or interact with charged particles, the released photon energy would be of the order of $10^{21}\eV$ or beyond if one recalls energy estimates of the previous section.
  Furthermore, if the accelerated monopole decays, it will produce photons of sufficiently high energy, which in turn can produce particle-antiparticle (e.g., proton, anti-proton) pairs.  
It seems not unlikely that the emission of UHE photons from monopoles accelerated to extremely high energy could occur near the Earth and be observed.
The energy of a monopole could be significantly reduced just as a result of the emission of UHE photons even before reaching  the Earth's atmosphere, and perhaps even at very high altitudes (e.g., in the van Allen belts), which would completely change the picture when it comes to detection and identification of monopoles. Then a set of atmospheric showers could disperse perhaps over a big area (depending on the photon emission place).  Even in the absence of detection, observational limits on monopole properties could be set.
 
Besides the acceleration processes, as sources of UHE photons, one can also consider massive monopolia \cite{hill1983}.
A bound state of a magnetic monopole and its anti-partner, called monopolium, may be formed naturally in the Universe and be easier to find than single monopoles. 
If massive enough, 
 the relict GUT monopolia with masses of $10^{16}\GeV$ produced in the very young universe
 may have survived up to the present
 in ultra-long-lived metastable states \cite{hill1983}. The monopolia would then decay by emitting radiation in the form of gluons, Z-bosons and photons, releasing a total energy of $2{\times}10^{16}\GeV$ in less than $10^{-38}\,{\rm sec}$ in the final annihilation stages. Such events, apart form the multitude of particles visible as cosmic rays, may produce high-energy gamma radiation. The number of various gluons, Z-bosons and photons produced in such an event in a given energy window can be calculated (taking into account fragmentation to secondary products). {\bf One can still expect on average several UHE photons with energy of $10^{19}\eV$ or more}, as follows from diagrams 2 and 3 in  \cite{hill1983}. Due to strong coupling between photons and monopoles one can also consider multi-photon annihilation process of monopole--antimonopole pairs
or their  bounded state -- monopolium -- into  $2n$ photons. Such decays were investigated in  \cite{fanchiotti2017}.

These short considerations above allow us to draw the conclusion that,  if magnetic monopoles are present in the Universe, not only should UHE values be accessible  across the Universe, but this energy may also be released in the form of UHE photons. Given this it is rather surprising to learn  that magnetic monopole searches so far have come up empty-handed. Nevertheless, gravitational radiation was predicted by General Relativity theory. This prediction was a mathematical construct following from the relativistic structure of the gravitation theory. Scientists did not give up their trust in this prediction in light of the theory's prior success, its consistency and aesthetic appeal. It took almost a century to reach sensitivity sufficient to detect gravitational waves. Would a similar scenario occur with magnetic monopoles?
 
\section{A magnetic monopole. What is it and why might it not exist?}

A magnetic monopole of Dirac is a structureless point-par\-ticle with undefined mass which resembles more a point-like electron than a composite proton. The monopole is described in the framework of Abelian Maxwell field theory in the presence of electrically charged quantum states.
The proton is not a purely electromagnetic particle, and its field-theoretical electrically charged models are considered.
The magnetic monopole might be realized in nature as an extended particle more similar to the proton than to the electron. 
Such particles can be described in terms of non-Abelian gauge field theory as magnetically charged solitons.
It is
important to understand the difference between these two kinds of magnetic monopoles.

\subsection{Dirac magnetic monopole}

The electric charge of isolated particles is always an integer multiple of elementary charge $e$, and there is no explanation for this quantization within the framework of the standard model of particles \cite{ellis1983}.
Particles quite unconnected with each other, such as 
 protons (a large composite 
particle) and electrons (a point-like particle), have equal absolute values of electric charge (defined in terms of Gauss integral formula over a
sphere of infinite radius) with the observational accuracy $10^{-21}$ \cite{bressi2011}.  In relativistic
quantum theory, the elementary charge corresponds to a pure number $\alpha{\equiv}\frac{e^2}{\hbar c}$ independent of units chosen
 ($\alpha^{-1}{=}137.035 999 084(21) $ is a currently recommended value \cite{CODATA}). 

In his work on {\it Quantized Singularities in the Electromagnetic Field}  \cite{dirac1931}, 
Dirac put forward an idea concerned with the reason for the existence of a smallest electric charge $e$, 
known to exist experimentally.
 Although it could be understood why electric charges of elementary particles can be mathematically equal (for example, if a magnetic pole exists, then electric charges must be integer multiples of some elementary charge, as follows from Dirac's formula), it is not known why $\alpha$ corresponding to the reference electron's charge has precisely this value and how it could be computed.  For a pure number of this kind one needs an explanation - this number must have 
to be named --  constructed algorithmically, say by identifying it with a unique convergent series, like it is for the $\pi$. It is known that the set of numbers that can be named are of measure zero in the set of all possible numbers 
on the real axis. It may be that there is not a mathematical structure that could be used to name $\alpha$;
then nobody will ever be able to compute $\alpha$
and one will have to relay on $\alpha$'s arbitrary experimental value. Dirac was looking for some explanation of this value. 
This problem remains still unsolved. According to Dirac, this problem is perhaps the most fundamental unsolved problem of physics. Dirac expressed his doubt whether any really big progress would be made in understanding the fundamentals of physics until this problem has been solved \cite{dirac1978}. 

Dirac starts his exposition of magnetic monopole theory \cite{dirac1931} 
with the observation that the phase of a normalized wave function $\psi$ can be globally changed by an arbitrary 
additive constant. The phase of $\psi$ at a particular point is thus not definite. Only the phase difference between any two points is definite. The difference need not be  
independent of the curve connecting these 
points unless the points are close enough to each other.  
A change in phase may occur around a closed curve. 
Dirac deduced from his analysis of various observables that,  
 in order for this not to give rise to ambiguity, the change in phase around any closed curve 
 must be the same for all $\psi$'s (leaving aside the role of representation). 
 This universality implies that 
 the phase increase between any 
 given pair of points along a particle's worldline  $x^{\mu}$ connecting these points will be determined solely
 by the ambient electromagnetic field $A_{\mu}$ and the worldline and thus through  
the line integral $\frac{e}{\hbar c}\int \!A_{\mu}\ud{x}^{\mu}$ evaluated along that worldline. Hence, the total
 increase in phase, $\Delta \Phi{\equiv}\frac{e}{\hbar c}\oint\! A_{\nu}\ud{x}^{\nu}$, around a closed path 
 will be nonzero for a non-integrable phase, 
 as already stated. In view of the Bohm--Aharonov effect \cite{aharonov1959}, it is clear that
only the phase factor $\exp{(\I\Delta \Phi)}$ and not $\Delta \Phi$ alone is physically meaningful \cite{wu1975}.  
  By Stokes' theorem applied to a continuously differentiable $A_{\mu}$, the $\Delta \Phi$ could be recast as a surface flux integral 
 $\frac{e}{2\hbar c}\iint F_{\mu\nu}\ud{x}^{\mu}\wedge \ud{x}^{\nu}$ evaluated over any surface stretched 
 across the closed curve. 
For a genuine Maxwell field, the $\Delta \Phi$ will vanish  
 in the limit when the closed curve gets shrunk to a point, even if the integration surface 
were to close and remain finite in that limit.\footnote{By Stokes' theorem, 
the discussed flux integral over a closed surface can be recast as a volume integral 
$\iiint\! \partial_{\alpha}F_{\mu\nu}\ud{x}^{\alpha}\wedge \ud{x}^{\mu}\wedge \ud{x}^{\nu}{\equiv}0$ 
identically vanishing on account of  
the structural identity $\partial_{\alpha}F_{\mu\nu}{+}\partial_{\mu}F_{\nu\alpha}{+}\partial_{\nu}F_{\alpha\mu}{\equiv}0$ satisfied by Maxwell fields. } 
However, for a closed spatial surface $\Sigma$ as perceived in some inertial reference frame, 
the surface integral
defines the total magnetic flux through 
$\Sigma$ (which must be zero).  

Therefore, in order to stay within the Maxwell theory framework and allow for a nonzero magnetic charge, 
it is necessary that  $A_{\mu}$ be somehow
 singular.
  The singularity might reside in a gradient contribution
 $\kappa_{\mu}{\equiv}\partial_{\mu} \kappa$ from a function $\kappa$ for which the integrability condition 
 requiring $\partial_{\nu}\kappa_{\mu}{=} \partial_{\mu}\kappa_{\nu}$ everywhere (equivalent here to Schwarz' theorem
 for a twice continuously differentiable function) is violated.\footnote{
 The classical example of such a singular function is the angular phase $\phi{=}\arctan(y/x)$ of a complex function $\exp{(\I\phi)}$ considered as a function on a plane: the line integral of the gradient field $[\partial_x\phi,\partial_y\phi]$ along a circle of any radius centered at $x{=}y{=}0$  is $2\pi$, whereas the corresponding `curl' field $\partial_xw_y-\partial_yw_x$ vanishes outside the center and is not defined there in terms of the 
 classical definition of derivatives. To interpret this in terms of Stokes' theorem on the plane, one has either to assume that the plane is punctured at the center and there is no field in that plane, or that the plane is smooth and the curl field is a distribution supported entirely at the center.}  This singularity could arise because of considering  a qualitatively more complicated system than merely the field alone, that is, a quantum particle interacting with a field.  
 Dirac showed that non-integrable derivatives $\kappa_{\mu}$ such as those originating from the indefinite quantum phase component $\kappa$, are of this kind and can be consistently reinterpreted in terms of electromagnetic field. 
 
 Namely, Dirac identified the $\kappa$ responsible for the change in phase around any closed curve 
 by noticing that any wave-function $\psi$
  can be represented as a product  $\psi{=}\tilde{\psi} \exp(\I \kappa)$. Here, 
 $\tilde{\psi}$ is normalized and has a definite phase, while  $\exp(\I \kappa)$
   involves 
  the indefinite part of the phase, however, with definite derivatives $\kappa_{\mu}{\equiv}\partial_{\mu}\kappa$.\footnote{Earlier, it was assumed that the phase difference between two points is definite, and therefore, this should hold in the limit of infinitesimally distant points.}       
 Furthermore, for all wave functions and their linear combinations with constant coefficients to acquire the same 
  change  $\Delta \Phi$ along a given closed path, it suffices that $\kappa_{\mu}$'s of different $\psi$'s will differ  between each other by the gradient 
  of a smooth function (this means that it is allowed to change $\kappa$ in each instance of $\psi$ by adding an arbitrary smooth function).  Now, one can suppose that $\psi$ is a state of a particle in free motion for which the indefiniteness of phase occurs. Acting with the quantum momentum  operator $\hat{p}_{\mu}{=}\I\hbar\partial_{\mu}$, 
  one can see that $ ({\rm e}^{\I\kappa} \hat{p}_{\mu} {\rm e}^{{-}\I\kappa})\psi{=}(\hat{p}_{\mu}{+}\hbar \kappa_{\mu})\psi$, which means that the classical four-momenta $p_{\mu}$ and $\tilde{p}_{\mu}$ corresponding to the respective states 
  $\psi$ and $\tilde{\psi}$ are related by $\tilde{p}_{\mu}{=}p_{\mu}{+}\hbar \kappa_{\mu}$, and so the definite phase state $\tilde{\psi}$ corresponds to the motion of charge $-e$ in the electromagnetic potential
  $A_{\mu}{=}{-}\frac{\hbar c}{e}\kappa_{\mu}$.  Since for an indefinite phase the integrability condition $\partial_{\nu}\kappa_{\mu}{=}\partial_{\mu}\kappa_{\nu}$ is violated somewhere, there must be an effective electromagnetic field $F_{\mu\nu}$ present.
A similar calculation for a state $\psi$ in a non-vanishing field leads to gauge transformation relations between momenta and between potentials in agreement with Weyl's Principle of Gauge Invariance \cite{weyl1929}.  This is the way that Dirac consistently incorporated indefiniteness of phase into the framework of ordinary quantum theory of charged particles in the presence of Maxwell fields.  This can be regarded as the first essential step in the construction of magnetic monopoles by  Dirac.

The second essential step in Dirac's construction starts with noticing that a wave function $\psi$ is not changed when
its phase gets increased by $2\pi n$, with $n$ being an integer. 
This implies that the difference between total increments in the phases of different wave functions around a loop will also be integer multiples of $2\pi$ in general, including the special case discussed earlier when $n{=}0$. 
However, by the continuity of wave functions as solutions of wave equations, this is not so when a loop 
is sufficiently small, in which case the total increment is only due to the encompassed flux, which also must be  small and vanishing in the limit of the stretched surface (and thus also the loop) being shrunk to a point.
However, there is one exception possible at nodal points where a given $\psi$ vanishes, and so the phase of $\psi$ cannot be defined.
The locus of nodal points forms a line as perceived in space, because a vanishing complex valued function provides two independent scalar constraints involving four spacetime coordinates. 
Hence, the total change in phase around a small loop
encircling the nodal line of $\psi$ will be equal to an integer multiple of $2\pi$ (depending on the particular $\psi$ considered) plus a universal contribution from the flux through a surface stretched on that small loop. 
The flux contribution in this case is universal for all $\psi$'s, irrespective of whether
a particular $\psi$ has or has not a nodal line passing through the considered surface. 
 Then, Dirac recalls the standard 
network of closed small loops argument to conclude that this result will be similar for
any large closed curve -- now the nodal line of a continuous $\psi$ can pass several times through a large 
 surface stretched on the closed curve, therefore the integer multiple of $2\pi$, now being a sum $2\pi\Sigma_i  n_i$, takes into account the respective positive and negative
 integer contributions, while the flux integral is now taken over that large surface. The final step is to close 
 the considered surface and to see that for any closed surface $\Sigma$ -- that is, one without a boundary closed curve -- the sum $2\pi\Sigma_i  n_i$ must be proportional to the
 total magnetic flux $4\pi\Phi_{B}$ through  $\Sigma$ (with a proportionality factor implied by the formula defining the contribution from the non-integrable phase $\kappa$ to the electromagnetic potential $A_{\mu}$, as discussed earlier), namely
$$2\pi\Sigma_i  n_i{+}\frac{e}{\hbar c}\cdot 4\pi \Phi_B{=}0,\qquad \Phi_B{\equiv}\frac{1}{4\pi}\cdot\frac{1}{2}\oiint_{\Sigma} F_{\mu\nu}\ud{x}^{\mu}\wedge \ud{x}^{\nu}.$$ 
The $\Phi_B$ is by definition the total magnetic charge enclosed within $\Sigma$.
 Since the flux in a given field must be the same irrespective of the particular $\psi$ being considered in that field, the result of the summation in $\Sigma_i  n_i$ must be universal too (although the set of integers $n_i$ may be different for different $\psi$'s) and so it can be denoted by some unique integer.
 If the integer is nonzero, it means that the nodal line of a given $\psi$ ends at some point inside the considered closed surface. Moreover, since the closed surface can be made infinitesimally small, one can infer that this point must be the point of origin for nodal lines of all $\psi$'s (although the lines may be different for different $\psi$'s).\footnote{It can be recalled, that the nodal line of a $\psi$ is something different from the singularity 
line of the part of $A_{\mu}$ connected with the non-integrability of the phase of $\psi$; while the nodal line of a $\psi$ is 
physically objective (phase independent) characteristic of a $\psi$ in a given field,  the latter singularity line can be deformed freely by means of smooth gauge transformations such that the endpoint of the singularity line remains fixed at the monopole position.}
At the position of the point, there is a magnetic monopole present with magnetic charge $g_n$ 
connected with the electric charge $e$ through Dirac's simple formula 
\begin{equation}g_n{=}ng,     \qquad g{=}\frac{\hbar c}{2e},\quad n{=}0,\pm1,\pm2,\dots\label{eq:Dirac_formula}\end{equation}
 with an integer $n$ characterizing the nodal lines ending at the monopole position.
 There is opposite singular magnetic flux along the string-like nodal line, compensating the flux of the magnetic charge through the spherical surface encompassing that charge. The singular flux could be potentially detected by means of the Bohm--Aharonov effect. However, since the corresponding change in phase is  an integer multiple of $2\pi$, it has no effect on a quantum particle in the field of the string.   The above relation between the smallest magnetic and electric charges $g$ and $e$ ensures that the string attached to the magnetic monopole is unobservable.  Therefore, the Dirac monopole acts as a genuine magnetic charge.
 
 The presence of the nodal line (called Dirac string)  emanating from the magnetic monopole position is associated with 
a singularity line of the gauge potential. In Cartesian coordinates in Minkowski spacetime, the static field of a Dirac magnetic monopole can be described by
the gauge potential $A{=}g\frac{x \ud{y}{-}y \ud{x}}{r(r{+}z)}$ (with $r{=}\sqrt{x^2{+}y^2{+}z^2}$ and to the extent of a constant dimensional factor),  which is 
singular on a straight semi-line $z{=}{-}r{\leqslant}0$ ($x{=}0{=}y$).    The corresponding gauge-invariant field $F{\equiv} d A$ reads \\$F{=}g\frac{x \ud{y}\wedge\ud{z}{+}y \ud{z}\wedge\ud{x}{+}z\ud{x}\wedge\ud{y}}{r^3}$ and is singular only at the center $r{=}0$ (the position of the magnetic monopole). The form $A$ is indeterminate up to the total derivative of a scalar field. Another possible gauge transformed form, $A'{=}A{-}2g d\phi{=}{-}g\frac{x \ud{y}{-}y \ud{x}}{r(r{-}z)}$, where
$\phi{=}\arctan(y/x)$, is singular on the straight semi-line $r{=}z{\geqslant}0$ ($x{=}0{=}y$), while $F$ remains the same (note that the scalar field used to generate the gauge transformation is not smooth). It is seen that the singularity line of the gauge potential could be altered  by means of singular gauge transformations
such that the end-point stays fixed at the center, and so the singularity line of the gauge potential has no physical meaning.
 
The elementary flux of a magnetic monopole can be obtained in a more straightforward way using the non-integrability of the angular variable. This was presented 
on the occasion of introducing  Dirac's electric monopole \cite{AStar1984} whose charge is determined by means of improper gauge transformation on a null plane. Although the electromagnetic potential $A_{\mu}$ is not gauge-invariant, the sum $\frac{e}{c}A_{\mu}{+}\partial_{\mu}S$, where $S$ is the phase of a quantum charged system, is gauge invariant: $\frac{e}{c}\delta A_{\mu}{+}\partial_{\mu}\delta S$ ${=} 0$, where the $\delta$ symbol denotes the change acquired after performing a gauge transformation (not necessarily infinitesimal). Now, performing an improper gauge transformation ${-}\delta S/\hbar{=}\phi{\equiv} \arctan{y/x}$, which leads to the allowable $2\pi\hbar$ increase in the phase round a closed loop (here, encircling the $z$ axis), it follows that $\frac{e}{\hbar c}\delta A_{\mu}{=}\partial_{\mu}\phi$, and hence $g{\equiv}\frac{1}{4\pi}\oint \delta A_{\mu}\ud{x}^{\mu}{=}$ $\frac{\hbar c}{2e}$ as expected for the elementary magnetic charge.

There is also an interesting result due to Jackiw \cite{jackiw1985} which shows that it is feasible to introduce the Dirac magnetic monopole in a gauge-invariant manner without the singular electromagnetic four-vector potential.  Jackiw's construction relays on the 
non-commutativity of kinematical momenta in the Lorentz--Heisenberg system defined by brackets: $[r^i,r^j]{=}0$, $[r^i,\pi^j]{=}\I\hbar\delta^{ij}$, $[\pi^i,\pi^j]{=}\I\frac{e\hbar}{c}B^{k}\epsilon_{k}^{\phantom{k}ij}$ and a Hamiltonian
$H{=}$ $\frac{1}{2m}{\pi^i\pi_i}$. The corresponding equations of motion make sense for  any (not necessarily source-free) magnetic vector $B^i$, however, with violated Jacobi identity $\epsilon_{ijk}[\pi^i,[\pi^j,\pi^k]]{=}\frac{2e\hbar^2}{c}\partial_k B^k{\neq}0$ implying non-asso\-cia\-ti\-vi\-ty in the composition of three finite translations in the presence of magnetic sources. To regain associativity required by the usual quantum formalism, a condition must be imposed on the total flux of magnetic field emerging out of a tetrahedron formed from a composition of the arbitrary three translation vectors.   The condition appears equivalent to Dirac's formula for magnetic charge \eqref{eq:Dirac_formula}. Importantly, Jackiw's construction demonstrates that {\it quantal magnetic sources must be structureless point particles} \cite{jackiw2003}. This observation points to some essential difference between elementary Dirac's monopole and field-theoretical realizations of magnetic monopoles discussed in the next subsection.

Finally it should be remarked that using Dirac formula one can predict 
 a much stronger coupling constant for a magnetic monopole--photon system than that for  electron--photon system. 
Given an electric charge in accelerated motion, various radiation formulas  are obtained in terms of 
Li{\'e}nard-Wiechert potential method and then improved by QED corrections. 
   According to Dirac formula \ref{eq:Dirac_formula}, the interaction force between two quanta of 
 magnetic monopoles is ${\sim} 4692$ times stronger than the interaction between two elementary electric charges. Hence, the emission amplitudes in the leading order are expected to 
be $(\frac{g}{e})^2{=}(n/2\alpha)^2{\sim}4692n^2$ times greater in the case of a magnetic charge than analogous amplitudes for electric charges.
Roughly speaking, this conclusion is reached by recoursing to a symmetry of Maxwell equations, 
thus replacing  in the formulas the electric charge  with magnetic charge  and the 
 Faraday tensor $F_{\mu\nu}$ with its dual $\frac{1}{2}\epsilon_{\mu\nu\alpha\beta}F^{\mu\nu}$ 
 (modulo some subtleties not important here);  see \cite{dooher1971}. 
 
\subsection{Field-theoretical realization of elementary magnetic monopoles}

The purpose of this section is to illustrate the difference between  the elementary point-like magnetic monopole of Dirac  and topological magnetic monopoles which are field-theoretical realizations of the elementary monopole.

  To obtain a field-theoretical realization of a Dirac magnetic monopole, a general unifying compact gauge group is chosen. It is such that it could be spontaneously broken down to the symmetry group of the Standard Model
        $U(3){\times}U(2){\times}U(1)$ so that the  electromagnetic field could emerge. The $SU(5)$ grand unified theory \cite{georgi1974} is an example in which monopoles arise \cite{scott1980}.  Topologically nontrivial solutions are interpreted as magnetic monopoles, whose charge appropriately identified with the topological charge of a given finite-energy field configuration becomes naturally quantized \cite{weinberg1996}. Such solutions are static non-point-like and  massive soliton lumps with finite width scale and the matter field quickly fades away with the distance. Asymptotically, only the part of the system is dominant which is mathematically equivalent to the magnetic field of an isolated point pole but without a string 
characteristic of Dirac monopole.                  
                  Predicted masses of magnetic monopoles within such theories vary by orders of magnitude and
can be as high as $10^{16}\GeV$ for super-massive monopoles, which might have appeared in high abundance in the early Universe before or after the inflation epoque and survive to the present \cite{preskill1984,weinberg1996}.

In 1974, 't~Hooft  \cite{hooft1974} and Polyakov \cite{polyakov1974}
found a stable monopole solution in a Yang-Mills-Higgs system with an $SO(3)$ gauge group and an isovector field, which is the simplest model in which magnetic monopoles appear (see \cite{witten1979}). Their construction (which  avoids the introduction of Dirac's string) could be repeated for
any group-valued gauge field for which the ordinary $U(1)$ gauge as a subgroup  can be defined equivalent to the gauge symmetry group of electromagnetism (see, for example, a construction for general non-commutative gauge fields \cite{wu1975}). The construction will be sketched below for completeness.

One starts with a gauge fields theory with some matter field forming a closed coupled system described by some Lagrangian density. The classic example in this respect considered is a non-Abelian model with a compact covering group because it has a spherically symmetric monopole as a solution. The model belongs to the class of Georgi-Glashow non-Abelian models  introduced to describe hierarchy of particles' masses emerging under spontaneously broken gauge symmetry  \cite{georgi1972}.  Assuming $SO(3)$  local  gauge group  for the illustration  in the simplest case,
  the respective Lagrangian density is a linear combination of three
parts:  the Yang-Mills term ${\propto} F^a_{\mu\nu}F_a^{\mu\nu}$, 
  the matter field term ${\propto} D_{\mu}\Psi_aD^{\mu}\Psi^a$ (involving coupling to Yang-Mills fields through 
  a covariant derivative $D$), and the potential term for a triplet of matter scalar fields ${\propto}\lambda(\Psi_a\Psi^a{-}\rho^2)^2$. Here, the summation is understood with respect to both kinds of indices separately, and $\rho,\lambda $ are fixed constants. 
 The covariant derivative acting on $\Psi^a$ is defined as $D_{\mu}\Psi^a{=}\partial_{\mu}\Psi^a{+}
\tilde{g}\epsilon^a_{\phantom{a}bc}A^b_{\mu}\Psi^c$,  while $F^a_{\mu\nu}{=}\partial_{\mu}A^a_{\nu}{-}
 \partial_{\nu}A^a_{\mu}{+}\tilde{g}\epsilon^a_{\phantom{a}bc}A^b_{\mu}A^c_{\nu}$. Here,  $A^a$ is the connection field (gauge potential), $F^a$ is the associated curvature (gauge field strength) and $\Psi$ is a (vector-valued) isovector Higgs field.
  From these group-valued fields, one can construct a Maxwell-like field
 based on 't~Hooft \cite{hooft1974} definition $F_{\mu\nu}{\equiv} n_a(F^a_{\mu\nu}{-}\tilde{g}^{-1}\epsilon^a_{\phantom{a}bc}
 D_{\mu}n^bD_{\nu}n^c)$, invariant with respect to $SU(2)$ gauge symmetry.
 Now, the divergence of the dual field  
 $F^{\star\mu\nu}{\equiv}$ $\frac{1}{2}\epsilon^{\mu\nu\alpha\beta}F_{\alpha\beta}$ 
 (which would be vanishing identically for an ordinary Maxwell field) is non-va\-ni\-shing and reads
  $\frac{1}{2}\epsilon^{\mu\nu\alpha\beta}\partial_{\nu}F_{\alpha\beta}{=}{-}{4\pi}\tilde{g}^{-1}J^{\mu}$
  where \\
  $J^{\mu}{\equiv}\frac{1}{8\pi}\epsilon^{\mu\nu\alpha\beta}\epsilon_{abc}\partial_{\nu}n^a
  \partial_{\alpha}n^b\partial_{\beta}n^c$ is a conserved topological
  current, $\partial_{\mu}J^{\mu}$ ${=}0$,  
  here $n^{a}{\equiv} $ $\frac{\Psi^a}{\sqrt{\Psi^b\Psi^b}}$ is a unit director of the Higgs field.  With this current, one can associate a topological charge
  that can be calculated, similarly as for ordinary electric charge, from a volume integral of the 
  time component of the current density: $Q{=}\int J^0\ud^3{x}$. In particular, in the static field case, the definition
  of $Q$ with the expression for $J^0$ substituted to the volume integral, in terms of topology, is simply the winding number -- a degree or index of the mapping $\vec{x}{\to} \vec{n}(\vec{x})$ between physical space and the domain of finite energy 
  static fields. Such defined $Q$
   attains only integer values. However, by analogy with Maxwell fields, we can write the $0$ component of the above 
   divergence of the dual static field as $\vec{\nabla}\vec{B}{=}4\pi \tilde{g}^{-1} J^0$, which upon integration gives
   the total flux of the magnetic field, and in this way, one obtains the associated magnetic charge $g_Q{=}Q\tilde{g}^{-1}$ of a soliton solution
   (in the static case $F_{0i}{=}0$. Thus, there is no corresponding electric field). 
      
   The lowest energy static field in the sector $Q{=}1$,  was found by 
    't~Hooft and Polyakov. The particular form of this solution is not important here. It suffices to say that the solution describes a  spatially extended finite energy stable soliton concentrated about the origin. The associated magnetic field of the 't~Hooft tensor $F_{\mu\nu}$ asymptotically overlaps with that of a point magnetic monopole. In effect  one can identify the solution as a monopole
with magnetic charge $\tilde{g}^{-1}$ and finite mass.  If the 't~Hooft tensor is to be interpreted as a genuine Maxwell field,  the coupling constant $\tilde{g}$ must be proportional to $e$. Now, the dimensional analysis shows that  one must have $\tilde{g}{=}\frac{e}{\hbar c}{=}\frac{\alpha}{e}$. Hence, the quantum of magnetic charge in this model is $g{=}\frac{e}{\alpha}$ which is twice the value obtained by Dirac. Additionally, in Schwinger's quantum field theory of electric and magnetic charges \cite{schwinger1966}, magnetic monopoles have charges twice the Dirac value. Schwinger made a comment on this discrepancy by saying that space-reflection considerations in his model require an infinite discontinuity line rather than the semi-infinite line appearing in Dirac's theory. He  gave several arguments to support his more restrictive value for the quantum of magnetic charge \cite{schwinger1966}. However, 't~Hooft points out that the Dirac quantization condition is fully restored in the Georgi--Glashow model when one considers isospin $1/2$ representations of the group $SU(2)$ \cite{hooft1974}.
        
 \section{Staruszkiewicz argument against the existence of magnetic monopole}

The argument against the existence of a magnetic monopole is rooted in the infrared or zero-frequency regime of electromagnetic fields. This regime is natural to consider since the information about electric and magnetic charges is encoded in the asymptotic part of the electromagnetic field that must decrease slowly 
enough in order to introduce a non-zero contribution to the Gauss flux integral (large distances mean low frequencies). By solving the bremsstrahlung radiation field, it also becomes clear that asymptotic electromagnetic fields, which are emitted when a
charge $Q$ changes its four-velocity, are localized entirely outside the light cone. 
It is therefore clear that spatial infinity must play the central role in the quantum theory of the electric charge. 
Staruszkiewicz formulated  such a theory \cite{AStar1989a}, and its structure turns out to depend in a nontrivial way on the numerical value of the fine structure constant $\alpha$. 

As an important part of his theory, 
Staruszkiewicz gave a simple argument against the 
existence of magnetic monopoles.
The argument was first proposed in an essay in honour of Yakir Aharonov \cite{AStar1998a}, then mentioned in \cite{AStar1997,AStar1998b,AStar2002a}.
The argument will be described in what follows. However, first one has to understand the residual structure
of the electromagnetic field at the spatial infinity that carries the information about electric and magnetic charges.
The structure of such fields has been described
by Alexander and Bergmann \cite{alexander1984}. It turns out 
that this structure can be equivalently described by two independent noninteracting scalar massless fields living on 2+1 dimensional deSitter spacetime. 

\subsection{Spatial infinity}
The outer part of the light cone (sufficient to investigate the spatial infinity) can be covered with the (hyper)spherical coordinates $$x^0{=}\chi \sinh{(\psi)},\qquad
\vec{x}{=}\chi \cosh{(\psi)}\cdot\vec{n}(\theta,\phi),$$ with  $\chi$ being the spatial 
coordinate ($0{<}\chi{<}{+}\infty$), $\psi$ the 
time-like coordinate (${-}\infty{<}\psi{<}{+}\infty$), and  a unit vector  $\vec{n}$ is
parameterized with ordinary spherical angles $0{\leqslant}\theta{\leqslant}\pi$, $0{\leqslant}\phi{<}2\pi$.
A time-like  de Sitter hyperboloid 
$\vec{x}^2{-}(x^0)^2{=}\chi^2$ (with $\chi$ regarded as a constant parameter) is a simple-connected three-dimensional curved spacetime with constant Ricci curvature $R{=}6/\chi^2$.
For considering asymptotic fields (fields in the limit $\chi{\to}{+}\infty$) or scale-independent fields,  one can instead consider the unit hyperboloid 
$\vec{x}^2{-}(x^0)^2{=}1$ as representing spatial infinity outside 
the light cone \cite{sommers1978}. 
The intrinsic metric on the unit de Sitter hyperboloid is diagonal when expressed in the adopted coordinates:
$$g_{ik}{\equiv}\chi^{-2}\eta(\partial_ix,\partial_jx){=}{\rm diag}[{-}1,\cosh^2(\psi),\cosh^2(\psi)\sin^2\theta],\quad i,j{=}1,2,3.$$  
It can be regarded as defining the metric of the spatial infinity.
It is seen that spatial infinity is homeomorphic and 
isometric to the unit hyperboloid \cite{alexander1984}.
 The coordinate $\psi$ plays the role of time, while 
spaces of constant time are unit two-spheres parameterized with $\theta,\phi$.
 The spatial infinity is the spacetime scene for asymptotic Maxwell fields and for their quantization.
It is important to observe that spatial infinity has a well-defined Cauchy surface, which is a feature desirable 
in the quantum field theoretical context.

\subsection{Zero frequency electromagnetic field and electron's charge as a quantum object}

The information on charges are encoded in the asymptotic part of the electromagnetic field that decreases slowly 
enough (this is reminiscent of the Gauss integral formula for the total charge -- as one can say,  
the electric charge resides at spatial infinity \cite{AStar1998b}). Described in terms of the 
electromagnetic gauge potential $A^{\mu}(x)$, this part is homogeneous of degree ${-}1$, depending on the 
direction of  the radius vector $x$ in a Lorentzian reference frame distinguished by the homogeneity property. 
Hence, the asymptotic field is effectively a function defined on the deSitter hyperboloid described earlier.
The asymptotic part can be extracted from any field by taking the Gervais--Zwanziger limit 
$A_{\mu}(x){\to} \lim_{\lambda{\to}{+}\infty}\lambda A_{\mu}(\lambda x)$ \cite{gervais1980} (leaving invariant  
fields $A_{\mu}$ homogeneous of degree ${-}1$).   Fields $A_{\mu}$ of this homogeneity  still exhibit a residual gauge freedom: 
$A_{\mu}{\to} A_{\mu}{+}\partial_{\mu}\Phi$, where $\Phi$ is a function homogeneous of degree $0$, since then 
$x^{\mu}\partial_{\mu}\Phi(x){=}0$ as follows from the Euler theorem on homogeneous functions.  
The asymptotic free fields satisfying the wave equation $\square A_{\mu}{=}0$ can be called zero frequency fields. 
One can see this by performing the Gervais--Zwanziger limit on a representation of $A_{\mu}$
as an integral of Fourier amplitudes 
over some invariant volume
on the light cone of null wave-vectors.
In this limit, the integral reduces to an integral of the limiting Fourier amplitudes 
effectively localized on the tip of the light cone.
In terms of the field tensor $F_{\mu\nu}$, if the limit $\lim_{\lambda\to{+}\infty}\lambda^2F_{\mu\nu}(\lambda x)$ exists and is non-zero, the field has a zero-frequency part; otherwise, if the limit is vanishing, there is no zero-frequency part in $F_{\mu\nu}$.
It should be noted that being 
a zero-frequency field is a Lorentz-invariant feature, while a non-zero frequency wave can be Doppler-shifted to an 
arbitrary frequency wave. These two regimes of electromagnetic radiation are qualitatively different.

  When an electric charge is scattered, the radiation field can be complicated. However, the zero-frequency asymptotic part of the field, 
  which can be extracted from the total radiation field by taking the Ger\-vais--Zwanziger limit, is quite simple.  It is equivalent to one emitted by an electric charge that abruptly changes its four-velocity at the point of scattering in the origin of the coordinate system, and which before and after being scattered remains in inertial motion, respectively, with four-velocity $u^{\mu}$ and   $w^{\mu}$.  Finding the radiation field emitted in such a process is a standard problem in classical electrodynamics. The result is such that the field vanishes both in the past light cone and in the future light cone and is non-vanishing
only outside the light cone. The radiation field can be described as a difference of two Coulomb fields, one at rest with respect to observer $u^{\mu}$ and the other at rest with respect to observer $w^{\mu}$. It is now clear that
  in any scattering process, the emitted zero-frequency field is universal and determined only by the initial and the final four-velocity, $u^{\mu}$ at $t{=}{-}\infty$ and   $w^{\mu}$ at $t{=}{+}\infty$. It is now also clear that although the amount of time available for the zero-frequency field at the spatial infinity is infinite, it is limited by the opening of the light cone, $|x^0|{<}|\vec{x}|$, because the zero-frequency radiation field is nonzero only outside the light-cone. 
  
  This property of a zero-frequency radiation field is perfectly tailored to understand why the elementary charge is a quantum object. Berestetskii, Lifshitz, and Pitaevskii criterion, \\
$\langle \vec{E}^2\rangle{\gg}{\hbar c}/(c\Delta t)^4$  \cite{BLP1982}, allows to decide if an electromagnetic field averaged over a time interval $\Delta t$ can be treated as classical. 
A static field would therefore be always classical as concluded in \cite{BLP1982}. This conclusion is paradoxical in view of the quantization of electric charges 
(a remark on this subject in a more philosophical context was made in \cite{AStar2002b} where the resolution of this paradox presented for the first time  in \cite{AStar1997}, is recalled). This conclusion is true, for example, when it comes to solving the Dirac equation for the hydrogen atom -- one considers quantum states of the electron in a classical Coulomb field of the proton, and one obtains good agreement of the predicted frequencies of emission lines with those observed. However, in the case of the zero-frequency field emitted in a scattering process, the criterion must be used carefully.   The radiation field in a given reference frame is a Coulomb field $q/r^2$ times  some kinematic factor which can be made to be of order $1$ by a suitable choice of the velocity change \cite{AStar1997}.
For a Coulomb field of charge $q{=}ne$: $\langle \vec{E}^2\rangle{=}n^2e^2/r^4$ and, by substituting $c\Delta t{=}2r$ to this inequality (which is the available averaging time at the spatial infinity from the opening of the light cone argument discussed earlier), it follows that the field will be classical if $n{\gg}\frac{\sqrt{\hbar c}}{4e}{\approx}\frac{1}{4}\sqrt{137}{\approx}2.93$. This observation was made in
  \cite{AStar1997}. This means that the Coulomb field ceases to be classical in the neighborhood of spatial infinity \cite{AStar2002b}. It means also that electric charges of the order of elementary charge $e$ (as determined upon application of the Gauss law) is not classical, in particular the electron's charge is a genuine quantum object. 
  If the value of fine structure constant was substantially larger one would not obtain such a sensible inequality.   
  With this observation it becomes clear how it is possible that proton's Coulomb field of the proton in the quantum theory of the hydrogen atom is purely  classical object while its amplitude determined by the proton's electric charge is quantized and involving Planck constant at the same time.
  
\subsection{Staruszkiewicz argument}

It is an experimentally established fact  that all electrically charged particles observed in nature are massive.
It means that electrically charged currents of these particles are confined to the future and past light cone, and 
quantum amplitudes are faded out exponentially at large distances by nonzero masses of the particles. From this, one can draw a physical conclusion
that the classical electromagnetic field $F_{\mu\nu}$ at the spatial infinity is totally free. 
Additionally, it must be homogeneous of degree ${-}2$ at spatial infinity if it is to carry an electric charge. Namely,
  $F_{\mu\nu}(\lambda x){=}\lambda^{-2}F_{\mu\nu}(x)$ for all positive $\lambda$ and for $(x^0)^2{-}\vec{x}.\vec{x}{\to}{-}\infty$.
Such  a field has two degrees of freedom described by scalar functions $\ef(x)$ and 
$\mf(x)$  satisfying the two equations
\begin{equation}\label{eq:emfuncts}
x^{\mu}F_{\mu\alpha}{=} \partial_{\alpha}[ {-}e x^{\mu}A_{\mu}(x)]{\equiv}\partial_{\alpha}\ef(x),
\quad \text{and}\quad 
\frac{1}{2}\epsilon_{\alpha\beta}^{\phantom{\alpha\beta}
\mu\nu}x^{\beta}F_{\mu\nu}{=}\partial_{\alpha}\mf(x).
\end{equation}
The first equation is proved in  \cite{AStar1998a} by a simple calculation assuming that $A_{\mu}(\lambda x){=}\lambda^{{-}1}A_{\mu}(x)$ 
for each positive
$\lambda$ and using the Euler theorem on homogeneous
 functions, here implying $x^{\nu}\partial_{\nu}A_{\mu}(x){=}{-}A_{\mu}(x)$.
 Besides the vector $x^{\mu}F_{\mu\nu}$, out of the field tensor and position vector, one can construct 
 in Minkowski spacetime another independent vector linear in the field: $\epsilon_{\alpha\beta}^{\phantom{\alpha\beta}
\mu\nu}x^{\beta}F_{\mu\nu}$ (more precisely, a pseudo-vector). Again, this vector can be shown to be a gradient. {This can be proved using the notation of calculus of antisymmetric forms. Consider a $1$-form $\omega{\equiv}\frac{1}{2}\epsilon_{\alpha\beta}^{\phantom{\alpha\beta}
\mu\nu}x^{\beta}F_{\mu\nu}\ud{x}^{\alpha}$. Then, the Maxwell equation $\partial_{\mu}F^{\mu\nu}{=}0$ and the homogeneity condition $x^{\alpha}\partial_{\alpha}F_{\mu\nu}{=}{-}2F_{\mu\nu}$ suffice to show that  ${\star}(\ud{\omega}){=}0$, where ${\star}$ denotes the Hodge star operator. This means that $0{=}{\star}(0){=}$ ${\star}({\star}(\ud{\omega})){=}{-}\ud{\omega}$. Thus, $\omega$ is an exact $1$-form; that is, there is a scalar $0$-form  $m$  such that $\omega{=}\ud{m}{\equiv} (\partial_{\mu}m )\ud{x}^{\mu}$.}
Furthermore, the divergence of left sides of the vector equations \eqref{eq:emfuncts} vanishes for arbitrary free Maxwell fields; hence,
scalars $\ef$ and $\mf$ must satisfy d'Alembert equations $\partial_{\mu}\partial^{\mu} \ef{=}0$ and $\partial_{\mu}\partial^{\mu} \mf{=}0$
and be homogeneous of degree $0$ (the latter property is seen directly from equations \eqref{eq:emfuncts}: on multiplying both sides of each equation by $x^{\alpha}$, the left hand sides are identically zero, while on the right hand sides one is left with the Euler scaling operator $x^{\alpha}\partial_{\alpha}$ acting on functions  $\ef$ and $\mf$). Equations \eqref{eq:emfuncts} together summarize  in the four-dimensional notation the structure of electromagnetic fields at spatial infinity \cite{AStar2002a}. The structure of these fields was described by Alexander and Bergmann \cite{alexander1984} who investigated electrodynamics at spatial infinity.

Once some solutions are given, the scalars $\ef$ and $\mf$ 
determine the physical field  $F_{\mu\nu}$ completely --  the vector equations \eqref{eq:emfuncts} can be solved for 
$F_{\mu\nu}(x)$ in a purely algebraic way. 
To give an example, one can consider  the  Coulomb field  of a point charge $e$ in inertial 
motion with four-velocity $u^{\mu}$. For this field,\footnote{Here, $(xx)$, $(xu)$, and $(uu)$ represent scalar products with the signature $({+},{-},{-},{-})$.}   $\ef(x){=}{e\,(ux)}/{\sqrt{(ux)^2{-}(xx)(uu)}}$ and 
$\mf(x){=}0$. 
The field, which is described by $A_{\mu}(x){=}e u_{\mu}\br{(ux)^2{-}(xx)(uu)}^{-1/2}$ in the 
Lorentz gauge $\partial^{\mu}A_{\mu}{=}0$, is homogeneous of degree ${-}1$, while the corresponding Faraday  
antisymmetric tensor, which is homogeneous of degree ${-}2$, reads $F_{\mu\nu}{=}e\br{ u_{\mu}x_{\nu}{-}u_{\nu}x_{\mu}}\br{(ux)^2{-}(xx)(uu)}^{{-}3/2}$.

For general fields  $F_{\mu\nu}(x)$ homogeneous of degree ${-}2$, scalars $\ef(x)$ and $\mf(x)$ (homogeneous of degree zero) can be regarded as
arbitrary 
functions defined over the unit de Sitter hyperboloid.  
They are effectively functions of $\psi$, $\theta$ and $\phi$ only (independent of $\chi$ and  satisfying the 
d'Alembert equations in this curved $2{+}1$D de Sitter spacetime) and thus 
are straightforwardly extendable to functions defined in whole outer part of the light cone. 
The Lagrangian density ${-}F_{\mu\nu}F^{\mu\nu}\ud^4{x}$ expressed in terms of such fields 
 becomes a difference 
of two identical Lagrangian densities \cite{AStar1989a,AStar1998a}
\footnote{
It is seen that $g^{ik}\partial_i\ef\partial_k\ef$ and $g^{ik}\partial_i\mf\partial_k\mf$ are 
both quadratic forms with signature $({+},{-},{-})$; thus, their difference is a quadratic form with signature $({+},{+},{+},{-},{-},{-})$
the same as the signature of arbitrary Maxwell Field $F_{01}^2{+}F_{02}^2{+}F_{03}^2{-}F_{23}^2{-}F_{31}^2{-}F_{12}^2$
 (in a given 
inertial frame, the latter form can be written as a difference $\vec{E}.\vec{E}{-}\vec{H}.\vec{H}$ with a contribution from 
electric field $\vec{E}$ and from magnetic field $\vec{H}$).}
\begin{equation}\label{eq:quadr}\!\!\!\!{-}F_{\mu\nu}F^{\mu\nu}\ud^4{x}{=}2\frac{\ud{\chi}}{\chi}\frac{\sin(\theta)}{\sech^2(\psi)}\ud{\psi}\ud{\theta}\ud{\phi}
\br{g^{ik}\partial_i\ef\partial_k\ef{-}g^{ik}\partial_i\mf\partial_k\mf}.\end{equation} Disregarding the factor 
$\ud{\chi}{/}\chi$, 
both Lagrangian densities are identical to one for a free massless scalar field on the $2{+}1$D de Sitter spacetime.  
The scalars $\ef$ and $\mf$ appear to be completely independent fields, Lorentz--invariantly separated from each other.  The action integral for this system regarded as confined  to the de Sitter unit hyperboloid $\mathcal{H}_1$ can be defined to be
$$S[\ef,\mf]{=}C\int_{\mathcal{H}_1}
\br{g^{ik}\partial_i\ef\partial_k\ef{-}g^{ik}\partial_i\mf\partial_k\mf}\sqrt{{-}g}\,\ud^3{\xi}.$$
Here, $g_{ik}$ is the metric tensor in arbitrary intrinsic coordinates $\xi^0,\xi^1,\xi^2$ on $\mathcal{H}_1$ (it is assumed that $g_{00}{>}0$);  $C$ is a positive dimensional constant introducing the correct absolute physical scale of the action integral.
The function $\ef$ is called the electric part, while the function $\mf$ is called the magnetic part of the field. In order to clarify the use of these names in simple terms, one may recall that in equation \eqref{eq:emfuncts} the electric function arises from Maxwell's tensor $F_{\mu\nu}$ and the magnetic function arises in the same way form the dual Maxwell tensor $\frac{1}{2}\epsilon_{\mu\nu}^{\phantom{\mu\nu}\alpha\beta}F_{\alpha\beta}$, in particular,  the magnetic function is vanishing for the Coulomb field considered earlier.  
It is evident that the fields  $\ef$ and $\mf$ enter the  Lagrangian density on quite the same footing.   Both the fields satisfy the free wave equation on  $\mathcal{H}_1$. The only difference lies in the opposite sign in front of the quadratic forms $g^{ik}\partial_i\ef\partial_k\ef$ and 
$g^{ik}\partial_i\mf\partial_k\mf$, and this difference, although
not harmless classically, is crucial 
when one wants to perform a quantization of that system.

Before proceeding further, a remark should be made concerning this sign difference. The overall sign 
of an action integral is not a matter of convention and cannot be changed arbitrarily. 
 It is known that the duality operation
is a symmetry of classical electromagnetism with electric and magnetic charges.   The symmetry is very useful in the practice of performing calculations  and making some predictions  in this theory. However, the sign of the Maxwell action integral gets reversed when magnetic fields are replaced with electric fields. 
As pointed out  by Hawking and Ross \cite{hawking1995} on the occasion of investigating electrically and magnetically charged black holes, the duality symmetry of the classical equations \emph{does not imply that it is a symmetry of the quantum theory, as the action is not invariant under duality}.\footnote{Semi-classical calculations leading to equal rates at which electrically and magnetically charged black holes are being created, made the authors to conclude that duality is a symmetry of the quantum theory, but in a very non-obvious way \cite{hawking1995}.  }
 If the sign of the action integral was not important there would be no essential difference between electric and magnetic functions  of the asymptotic electromagnetic fields, and  so $\ef$  and $\mf$ would be present in the theory on the same footing.

The difference in sign of the electric and magnetic parts mentioned above is essential and has physical consequences -- the wrong sign implies the existence of negative norm states not belonging to the framework of quantum mechanics.
Namely, upon quantization only one of the quadratic forms in the above action integral would lead to a positive definite inner product. The one with opposite 
definite inner product, introducing negative norm states, should be considered as non-physical and therefore should be rejected. 
A Lagrangian density with a correct 
sign is that which introduces a field with a positive kinetic term (that is, the one involving partial derivatives of the field  with respect to the time-like variable -- for example, $\psi$ in the discussed spherical coordinates on $\mathcal{H}_1$).
 It is seen that 
the kinetic term involving $\ef$ in the above action integral is positive, while    
the kinetic term involving $\mf$ is negative.
Therefore, Staruszkiewicz concludes that at the spatial infinity, the part of the electromagnetic 
field with the incorrect sign should be absent, which is achieved by putting $\mf{=}0$.
This statement can be rephrased by saying that magnetic monopoles should not exist.

As shown in \cite{AStar1998b}, this conclusion is not changed when the original Maxwell Lagrangian is extended by adding to 
${-}\frac{1}{16\pi}F_{\mu\nu}F^{\mu\nu}$ the CP-symmetry violating $\Theta$ term $\Theta \epsilon^{\alpha\beta\mu\nu}F_{\alpha\beta}F_{\mu\nu}$ involving the second independent invariant 
of the electromagnetic field. By adding the $\Theta$ term, one cannot change the signature of the quadratic form in \eqref{eq:quadr}. 
Indeed, the extended Lagrangian density, when expressed in terms of 
new electric and magnetic functions $\ef'$ and $\mf'$,
 attains the same quadratic form as in equation \eqref{eq:quadr} in which 
$\ef$ and $\mf$ have been replaced with $\ef'$ and $\mf'$ (the new and old sets of functions are related to each other as follows $\ef'{=}\ef \cosh{\upsilon}{-}\mf \sinh{\upsilon}$ and $\mf'{=}\ef \sinh{\upsilon}{+}\mf \cosh{\upsilon}$, where
$\sinh{(2\upsilon)}{=}8\pi\Theta$, one can see from this that the part with the negative kinetic term is increased with respect to the case without the $\Theta$ term). Therefore, the argument against magnetic monopoles still holds with the extended Lagrangian. Now, one should assume $\mf'{=}0$ to ensure that  upon quantization, the negative norm states do not arise. 
Witten showed that field-theoretical realizations of magnetic monopoles in CP-violating theories with the theta mechanism would not be 
electrically neutral and would carry even non-rational fractions of electric charge $Q{=}e(n{-}\frac{\vartheta}{2\pi})$, depending on vacuum angle parameter $\vartheta$ introduced by the $\Theta$ term \cite{witten1979}.
On the other hand,  in the framework of Staruszkiewicz theory \cite{AStar1989a}, it 
can be shown that electric charge must be quantized in units of electronic charge \cite{AStar1998a}.
Thus non-existence of magnetic monopoles is compatible with electric charge quantization \cite{AStar1998b}.
 
\section{Conclusions}

Ultra-high energy (UHE) photons with energies exceeding $10^{18}\eV$ could be detected
by Earth-based observatories. UHE photons are produced in various processes 
in which electrically charged particles take part.  
However, more exotic processes are also possible.
There are compelling theoretical premises in favor of the existence of isolated magnetic monopoles, especially when it comes to 
realisation of the monopoles in the framework of non-Abelian gauge fields. 
UHE photons could be produced in encounters of massive magnetically charged monopole-antimonopole pairs or in  processes associated with monopoles accelerated to high energies, typically $10^{21}\eV$ or beyond.
Observing UHE photons can pose constraints on the properties of magnetic monopoles. 
The predicted observational signatures of these particles are being sought in many dedicated experiments currently in operation. 
Currently, no magnetic monopoles have been found despite many attempts to detect them. 
Neither
intermediate mass monopoles below the GUT scale that could be produced in accelerators nor monopoles that should exist in the Universe and produce observable signatures have been observed. 
Monopole masses predicted by the field-theoretical realizations of magnetic monopoles 
are apparently too high, which might explain the negative experimental 
evidence for magnetic monopoles so far. However, there is another possibility.
There have been formulated arguments for why magnetic monopoles allowed by Dirac's theory 
might not be realized in nature \cite{herdegen1993,AStar1998a}. In particular, Staruszkiewicz's argument against magnetic monopoles is an important part of his quantum theory of the electric charge  \cite{AStar1989a}.
This argument invokes the  positivity of the norm in Hilbert space. This positivity is violated by quantum states of the  magnetic part of zero-frequency fields. If these arguments could not be refuted, one would have to conclude that, while isolated magnetically charged solitonic configurations can be considered as solutions in the standard model of particles, 
only magnetically neutral configurations of magnetic monopoles
 could be realized in nature. 

\bibliographystyle{ieeetr}
\bibliography{2022LBJJ_UHEphVerMgnMnpl_v2_ArXiv}

\end{document}